\documentclass[preprint,12pt,authoryear]{elsarticle}

\usepackage{amssymb}
\usepackage{amsmath}
\usepackage{amsthm}
\newtheorem{theorem}{Theorem}

\usepackage{xcolor}
\usepackage{comment}

\journal{Economics Letters}

\begin{document}

\begin{frontmatter}



\title{The Polarization Effect of Monopsonistic Lobbying} 

\author[label1]{Peter Shum}
\ead{petershum@ccu.edu.tw}
\ead[url]{https://hopanshum.com}
\cortext[cor1]{Corresponding author}
\affiliation[label1]{organization={National Chung Cheng University},
             state = {Chiayi},
             country={Taiwan}}



\begin{abstract}
Classical spatial models predict platform convergence, yet empirical polarization persists. This paper proposes a non-electoral mechanism: lobbying as a monopsonistic market for legislative support. Here, extreme benefactors must pay more to attract distant politicians, creating a rent gradient that rewards platform differentiation. We find that the unique equilibrium places politicians at $(\tfrac14,\tfrac34)$ for any monotone policy-production cost. Thus, polarization can arise solely from lobbying-market structure, independent of electoral incentives.
\end{abstract}
\begin{highlights}
\item Lobbying is modeled as a monopsonistic market for legislative support.
\item Monopsony power creates a rent gradient that rewards differentiation.
\item Equilibrium platforms diverge to $(1/4, 3/4)$ for any monotone cost.
\item Non-electoral incentives alone overturn the median voter prediction.
\end{highlights}

\begin{keyword}


Polarization \sep Lobbying \sep Monopsony \sep Spatial competition \sep Political economy
\end{keyword}

\end{frontmatter}



\section{Introduction}
\label{intro}
Classic spatial models \`a la \citet{Hotelling1929} and \citet{Downs1957} predict platform convergence to the median voter. Yet modern democracies exhibit persistent and intensifying polarization \citep{Yang2020}. Much of the existing literature has sought to explain these patterns through \textit{electoral} mechanisms such as primary competition, ideological shocks, strategic extremism, or valence asymmetries \citep{Glaeser2005,Matakos2016,Kishishita2022}.
\par
This electoral focus ignores a well-documented empirical reality: politicians devote the vast majority of their working hours to non-electoral activities. Time-use studies show that fundraising, legislative research, and interest-group engagement dominate the legislative calendar \citep{FouirnaiesHall2018,HallWayman1990,HallDeardorff2006}. Re-election rates remain exceptionally high \citep{GelmanKing1990,Lee2008,FowlerHall2014}, suggesting that holding office is a long-term career rather than a repeated contest. Politicians thus face incentives that extend well beyond the immediate demands of the ballot box.
\par
Lobbying drives this non-electoral activity. Rather than simple vote-buying, lobbying functions as a ``legislative subsidy'': providing the effort and expertise that reduce legislators' costs of engaging with policy \citep{HallDeardorff2006}. This aligns with evidence on the informational role of interest groups across sectors \citep[e.g.,][]{Kang2016,Bertrand2020}. We model this environment as a political labor market: benefactors demand legislative effort, and politicians supply it subject to differentiated policy-production costs.
\par
Empirical work directly links these incentives to polarization. Notably, \citet{Garlick2022} demonstrates that lobbying investments causally increase legislative party division. This establishes the need for a mechanism connecting the market for political support to spatial divergence.
\par
We focus on the demand for legislative effort. Some lobbying activities are highly concentrated; specific issues attract dense organized interest-group activity \citep{Baumgartner2009,Drutman2015}. These interests possess substantial market power relative to individual legislators. Furthermore, because the information provided functions as a non-excludable input, uniform compensation emerges as a natural outcome. \citet{dAspremont1979} famously established that \textit{monopoly} power drives differentiation on the seller side. We examine the analogous effect of \textit{monopsony} power on the buyer side, modeling the benefactor as a wage-setter for political support.
\par
This note develops a simple-yet-general model showing how monopsonistic lobbying can generate spatial polarization even when voters are unpolarized. Two politicians compete for a uniform wage offered by a single benefactor, who must pay the minimum amount required to attract both to support her preferred policy. Because this wage is pinned down by the more distant politician, extreme benefactors pay more, creating a rent gradient that rewards platform differentiation. The resulting equilibrium places the two politicians at $(\tfrac14,\tfrac34)$, demonstrating that the structure of the lobbying market alone can overturn the median-voter prediction.



\section{Model}
The ideology space is a continuum $[0,1]$. Two ex-ante identical politicians choose platforms $x_1,\, x_2 \in [0,1]$. A continuum of benefactors are uniformly distributed on this space. For each benefactor with ideology $b$, they aim to gather support from both politicians for a policy at $b$. Politician with an established platform $x_i$ incurs a utility cost of $c(|x_i-b|)$ to support the policy. The cost $c(\cdot)$ is strictly increasing in the distance $|x_i-b|$. This cost represents any political, informational, or reputational burden associated with backing a policy away from one's established platform $x_i$.
\par
\begin{figure}
  \centering
  \includegraphics[scale=1]{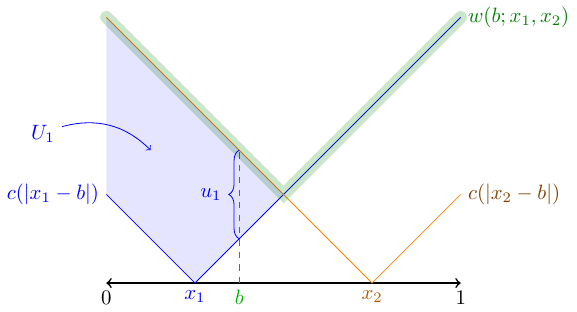}
  \caption[Illustration of the model setup.]{Illustration of the model setup with $x_1\neq x_2$ and $c(|x_i-b|) = |x_i - b|$. The cost and payoff of a politician with a platform $x_1$ is in {\color{blue}blue}. The cost of a politician with a platform $x_2$ is in {\color{orange!60!black}orange}. The wage benefactors pay is in {\color{green!60!black}green}, which is the V-shaped upper envelope of the costs.}
  \label{fig:1}
\end{figure}

A benefactor at $b$ offers lobbying investment, in which professional lobbyists convert into publicly provided service utility to the politicians. The benefactor must supply the minimum investment to induce both politicians to support the policy. Hence the induced service utility, effectively a uniform wage, is
\begin{equation}
  w(b;x_1,x_2) = \max \{c(|x_1-b|), c(|x_2-b|)\}
\end{equation}
An illustration of the cost and wage functions can be found in Figure \ref{fig:1}.
\par
The V-shaped wage upper envelope defines the core mechanism: centralist benefactors, being relatively close to both politicians, face low service costs, whereas extreme benefactors must pay a premium to attract the distant politician. In short, \textit{extremists pay more.} 
\par

\par
A politician at platform $x_i$ obtains interim utility
\begin{equation}
  u_i(b;x_1,x_2)=w(b;x_1,x_2)-c(|x_i-b|),
\end{equation}
which is nonnegative by construction since the benefactor chooses the wage to induce support from both politicians. Given the uniform distribution of benefactors on $[0,1]$, her interim expected utility is
\begin{equation}
  U_i(x_i,x_j)=\int_0^1 u_i(b;x_1,x_2)\,db.
\end{equation}

\section{Results}

\subsection{Benchmark}
To isolate the role of benefactors' monopsony power in controlling wages, consider a benchmark in which benefactors cannot adjust their lobbying investment to politicians' locations. In this scenario, politicians' policy-production costs are covered through alternative channels (such as publicly funded staffs), so the effective compensation takes the form of a competitive wage schedule $w_c$ that must be high enough to cover the highest possible costs across all $(x_1,x_2,b)$ tuples.

Since the competitive wage does not vary with platform choices, politicians face no strategic trade-off: their only consideration is to minimize their own policy-production cost. By the monotonicity of $c$, this cost is minimized at the center, so their best responses satisfy
\[
  x_1^c = x_2^c = \tfrac12.
\]
Thus the competitive benchmark features full convergence at the median, echoing the classical Hotelling prediction. This highlights that polarization cannot arise without the benefactor's ability to reduce wages when politicians move inward.

\subsection{Payoff Intuition}
Under monopsony, the benefactor at $b$ offers the minimum wage needed to induce both politicians to support the proposal. As illustrated in Figure~\ref{fig:1}, this wage is determined by whichever politician is farther from $b$. For the left politician at $x_1$, benefactors on the right never generate positive utility: for such $b$, the wage equals her own cost $c(x_1,b)$, leaving zero surplus. Positive utility arises only from nearby benefactors on the left, where the wage must be raised to compensate the right politician's larger cost. The situation is symmetric for the right politician.
\par
Hence each politician earns rents only from the side near her own platform. While moving inward would reduce a politician's own policy-production cost and, holding wages fixed, increase her interim utility, benefactors' monopsony power allows the distant benefactor to cut wages whenever this politician becomes cheaper to attract. In particular, under full centralization, benefactors can drive wages down to the point where politicians earn no rents. Thus full convergence to the center is not attractive: sustaining positive rents requires some degree of platform separation.
\subsection{Equilibrium}
We look for a pure-strategy perfect Bayesian equilibrium, interpreted as a profile $(x_1,x_2)$ in which each politician's platform is a best response to the other. The benefactor's wage schedule is given by $w(b;x_1,x_2)=\max\{c(|x_1-b|),c(|x_2-b|)\}$, and each politician chooses her platform to maximize her expected interim payoff before lobbies.
\begin{theorem}\label{thm:1}
With strictly increasing policy-production costs $c(|x_i-b|)$, the unique pure-strategy PBE (up to relabeling of players) is given by
\[
  x_1^\ast = \tfrac14, \quad x_2^\ast = \tfrac34.
\]
\end{theorem}
The proof is provided in \ref{app:proof}. The equilibrium reflects the rent-extraction logic described above: each politician positions herself to preserve a neighborhood of benefactors who must offer a high wage to attract the distant opponent.
\par
While the quarters profile $(1/4,3/4)$ is not maximal polarization, it yields substantial separation and aligns with the empirical patterns documented by \citet{Yang2020}, despite arising from a mechanism distinct from the theoretical account in their model.
\par
The robustness of the quarters equilibrium is notable when compared with classical spatial models such as \citet{dAspremont1979}, where the curvature of transportation costs is central to the equilibrium outcome. In the present model, by contrast, a politician has little influence over the wage in the neighborhood where she earns rents, since that wage is determined entirely by the distant opponent's cost. Her marginal incentives to adjust her platform can therefore be viewed in terms of the arc length of her own cost curve below the opponent's cost on each side. The best response is the point at which these arc lengths balance, a condition that depends only on the relative positions of $x_1$ and $x_2$ rather than on the curvature of $c(\cdot)$. Hence the equilibrium platforms remain $(\tfrac14,\tfrac34)$ for all strictly increasing cost functions.
\section{Discussion}
Our baseline model suggests several natural extensions. Allowing politicians to coordinate, or introducing more than two politicians with non-unanimity requirements, would alter the monopsonist's leverage and could change the extent of induced polarization. We leave such questions for future work.
\appendix
\section{Proof of Theorem \ref{thm:1}.}
\label{app:proof}

\begin{proof}
Fix $x_j \in [0,1]$. Politician $i$ chooses $x_i$ to maximize
\[
U_i(x_i,x_j)=\int_0^1 \left[w(b;x_i,x_j)-c(|x_i-b|)\right] db,
\]
where $w(b)=\max\{c(|x_i-b|),c(|x_j-b|)\}$.
\par
We compute the surplus regions. If $x_i \le x_j$, politician $i$ earns surplus only for
$b \le (x_i+x_j)/2$, where the wage is $w(b)=c(x_j-b)$. Thus,
\[
U_i(x_i,x_j)=\int_0^{(x_i+x_j)/2}[c(x_j-b)-c(|x_i-b|)]\,db.
\]
Similarly, if $x_i \ge x_j$, the surplus comes only from
$b \ge (x_i+x_j)/2$, where the wage is $c(b-x_j)$.
\par
Next, we compute the interior maximizers within the two surplus regions.
\par
\emph{Case A: $x_i \le x_j$.}  
Split the integral at $b=x_i$ to eliminate the absolute value, apply the Leibniz
rule, and use symmetry of $c$:
\[
\frac{\partial U_i}{\partial x_i}
= c\!\left(\frac{x_j-x_i}{2}\right)-c(x_i).
\]
The first-order condition along with the monotonicity of $c$ therefore yields the unique interior maximizer
\[
x_i^\dagger = \frac{x_j}{3}.
\]
\par
\emph{Case B: $x_i \ge x_j$.}  
A symmetric argument gives
\[
\frac{\partial U_i}{\partial x_i}
= c(1-x_i)-c\!\left(\frac{x_i-x_j}{2}\right),
\]
whose unique interior solution is
\[
x_i^\ddagger = \frac{2+x_j}{3}.
\]

Thus, the only candidates for best responses are $x_i^\dagger$ and
$x_i^\ddagger$. Next, we compare the candidates and find the best response to $x_j$. Define for $t\in[0,1/3]$ the function
\[
F(t)
:= \int_t^{3t} c(b)\,db - 2\int_0^t c(b)\,db.
\]
A direct substitution shows
\[
U_i(x_i^\dagger,x_j)=F\!\left(\frac{x_j}{3}\right),\qquad
U_i(x_i^\ddagger,x_j)=F\!\left(\frac{1-x_j}{3}\right).
\]

Differentiating,
\[
F'(t)=3c(3t)-c(t)>0
\]
by strict monotonicity of $c$, so $F$ is strictly increasing.

Therefore:
\[
U_i(x_i^\dagger,x_j)\ge U_i(x_i^\ddagger,x_j)
\quad\Longleftrightarrow\quad
\frac{x_j}{3}\ge \frac{1-x_j}{3}
\quad\Longleftrightarrow\quad
x_j\ge \frac12.
\]

Hence the best-response correspondence is
\[
BR_i(x_j)=
\begin{cases}
\{\tfrac{x_j}{3}\}, & \text{ if }x_j>\tfrac12,\\[0.3em]
\{\tfrac{2+x_j}{3}\}, & \text{ if } x_j<\tfrac12,\\[0.3em]
\{\tfrac{x_j}{3},\,\tfrac{2+x_j}{3}\}, & \text{ if }x_j=\tfrac12.
\end{cases}
\]
A pure-strategy equilibrium requires $x_i\in BR_i(x_j)$ for both players. Because one politician must lie below $1/2$ and the other above $1/2$, their best responses satisfy
\[
x_1=\frac{x_2}{3},\qquad
x_2=\frac{2+x_1}{3}.
\]
Solving yields
\[
(x_1^*,x_2^*)=\left(\tfrac14,\tfrac34\right).
\]

This is the unique equilibrium up to relabeling of politicians.
\end{proof}

\section*{Declaration of Generative AI-assisted technologies in the writing process}
During the preparation of this work, the author used Large Language Models in order to improve the readability and language of the abstract and introduction. The author reviewed and edited the content, and take full responsibility for the content of the publication.

\bibliographystyle{elsarticle-harv} 
\bibliography{polarization.bib}

@book{Baumgartner2009,
  title     = {Lobbying and Policy Change: Who Wins, Who Loses, and Why},
  author    = {Baumgartner, Frank R. and Berry, Jeffrey M. and Hojnacki, Marie and Kimball, David C. and Leech, Beth L.},
  publisher = {University of Chicago Press},
  address   = {Chicago},
  year      = {2009}
}

@article{Bertrand2020,
  title   = {Tax-Exempt Lobbying: Corporate Philanthropy as a Tool for Political Influence},
  author  = {Bertrand, Marianne and Bombardini, Matilde and Fisman, Raymond and Trebbi, Francesco},
  journal = {The American Economic Review},
  volume  = {110},
  number  = {7},
  pages   = {2065--2102},
  year    = {2020},
  doi     = {10.1257/aer.20180615}
}

@article{dAspremont1979,
  title   = {On Hotelling's ``Stability in Competition''},
  author  = {d'Aspremont, Claude and Gabszewicz, J. Jaskold and Thisse, Jacques-Fran\c{c}ois},
  journal = {Econometrica},
  volume  = {47},
  number  = {5},
  pages   = {1145--1150},
  year    = {1979}
}

@book{Downs1957,
  title     = {An Economic Theory of Democracy},
  author    = {Downs, Anthony},
  publisher = {Harper},
  address   = {New York},
  year      = {1957}
}

@book{Drutman2015,
  title     = {The Business of America is Lobbying},
  author    = {Drutman, Lee},
  publisher = {Oxford University Press},
  address   = {New York},
  year      = {2015}
}

@article{FouirnaiesHall2018,
  title   = {How Do Interest Groups Seek Access to Committees?},
  author  = {Fouirnaies, Alexander and Hall, Andrew B.},
  journal = {American Journal of Political Science},
  volume  = {62},
  number  = {1},
  pages   = {132--147},
  year    = {2018},
  doi     = {10.1111/ajps.12323}
}

@article{FowlerHall2014,
  title   = {Disentangling the Personal and Partisan Incumbency Advantages: Evidence from Close Elections and Term Limits},
  author  = {Fowler, Anthony and Hall, Andrew B.},
  journal = {Quarterly Journal of Political Science},
  volume  = {9},
  number  = {4},
  pages   = {501--531},
  year    = {2014},
  doi     = {10.1561/100.00014013}
}

@article{Garlick2022,
  title   = {Interest Group Lobbying and Partisan Polarization in the {United States}: 1999--2016},
  author  = {Garlick, Alexander},
  journal = {Political Science Research and Methods},
  volume  = {10},
  number  = {3},
  pages   = {488--506},
  year    = {2022}
}

@article{GelmanKing1990,
  title   = {Estimating Incumbency Advantage Without Bias},
  author  = {Gelman, Andrew and King, Gary},
  journal = {American Journal of Political Science},
  volume  = {34},
  number  = {4},
  pages   = {1142--1164},
  year    = {1990}
}

@article{Glaeser2005,
  title   = {Strategic Extremism: Why Republicans and Democrats Divide on Religious Values},
  author  = {Glaeser, Edward L. and Ponzetto, Giacomo A.M. and Shapiro, Jesse M.},
  journal = {The Quarterly Journal of Economics},
  volume  = {120},
  number  = {4},
  pages   = {1283--1330},
  year    = {2005}
}

@article{HallDeardorff2006,
  title   = {Lobbying as Legislative Subsidy},
  author  = {Hall, Richard L. and Deardorff, Alan V.},
  journal = {American Political Science Review},
  volume  = {100},
  number  = {1},
  pages   = {69--84},
  year    = {2006},
  doi     = {10.1017/S0003055406062010}
}

@article{HallWayman1990,
  title   = {Buying Time: Moneyed Interests and the Mobilization of Bias in Congressional Committees},
  author  = {Hall, Richard L. and Wayman, Frank W.},
  journal = {American Political Science Review},
  volume  = {84},
  number  = {3},
  pages   = {797--820},
  year    = {1990}
}

@article{Hotelling1929,
  title   = {Stability in Competition},
  author  = {Hotelling, Harold},
  journal = {The Economic Journal},
  volume  = {39},
  number  = {153},
  pages   = {41--57},
  year    = {1929},
  url     = {http://www.jstor.org/stable/2224214}
}

@article{Kang2016,
  title   = {Policy Influence and Private Returns from Lobbying in the Energy Sector},
  author  = {Kang, Karam},
  journal = {The Review of Economic Studies},
  volume  = {83},
  number  = {1},
  pages   = {269--305},
  year    = {2016},
  doi     = {10.1093/restud/rdv029}
}

@article{Kishishita2022,
  title   = {Do Supermajority Rules Really Deter Extremism? The Role of Electoral Competition},
  author  = {Kishishita, Daisuke and Yamagishi, Atsushi},
  journal = {Journal of Theoretical Politics},
  volume  = {34},
  number  = {1},
  pages   = {127--144},
  year    = {2022},
  doi     = {10.1177/09516298211061161}
}

@article{Lee2008,
  title   = {Randomized Experiments from Non-Random Selection in {U.S.} House Elections},
  author  = {Lee, David S.},
  journal = {Journal of Econometrics},
  volume  = {142},
  number  = {2},
  pages   = {675--697},
  year    = {2008},
  doi     = {10.1016/j.jeconom.2007.05.004}
}

@article{Matakos2016,
  title   = {Electoral Rule Disproportionality and Platform Polarization},
  author  = {Matakos, Konstantinos and Troumpounis, Orestis and Xefteris, Dimitrios},
  journal = {International Tax and Public Finance},
  volume  = {23},
  pages   = {1031--1055},
  year    = {2016}
}

@article{Yang2020,
  title   = {Why Are {U.S.} Parties So Polarized? A ``Satisficing'' Dynamical Model},
  author  = {Yang, Vicky Chuqiao and Abrams, Daniel M. and Kernell, Georgia and Motter, Adilson E.},
  journal = {SIAM Review},
  volume  = {62},
  number  = {3},
  pages   = {646--657},
  year    = {2020}
}



\end{document}